\documentclass[%
 reprint,
 amsmath,amssymb,
 aps,
]{revtex4-2}

\usepackage{graphicx}% Include figure files
\usepackage{dcolumn}% Align table columns on decimal point
\usepackage{bm}% bold math
\usepackage{xcolor}
\usepackage{xcolor}
\begin{document}

\preprint{}
%\begin{bibunit}

\title{\textbf{Interplay of Micellar Architecture and Viscosity Governs Active Droplet Motility} }% Force line breaks with \\

\author{Salini Kar,$^1$ Rohit V. Menon,$^1$ Sanbed Das,$^2$ Parth Pandya,$^3$ Sayantan Dutta,$^{2,*}$ Mithun Chowdhury$^{1}$}
%Lines break automatically or can be forced with \\
 
%\author{ }
\email{sayantan.dutta@iitb.ac.in}
\email{mithunc@iitb.ac.in}

\affiliation{$^1$Lab o\textit{f} Soft Interfaces (LoSI), Department of Metallurgical Engineering and Materials Science, Indian Institute of Technology Bombay, Mumbai 400076, India}

\affiliation{$^2$Computational Engineering Lab for Living Materials (CELL), Department of Chemical Engineering, Indian Institute of Technology Bombay, Mumbai 400076, India.}

\affiliation{$^3$Soft and Biological Matter Lab, Department of Physics, Indian Institute of Technology Kanpur, Kanpur 208016, India}

\date{\today}% It is always \today, today,
             %  but any date may be explicitly specified

\begin{abstract}
The autonomous motion of liquid crystal oil droplets in micellar media arises from spontaneous breaking of time-reversal symmetry via nonlinear coupling between Marangoni stresses and surfactant transport. While this phenomenon has been widely studied, the influence of micellar solute structure remains unexplored. By modifying micellar architecture using a structure-forming salt, we uncover a pronounced non-monotonic dependence of droplet velocity on salt concentration. Increasing salt simultaneously raises the medium viscosity and drives a transition of micelles from spherical to rod-like or worm-like morphologies. Using complementary experiments, we quantify the viscosity and micellar interaction lengthscale as functions of the salt-to-surfactant ratio and develop a theoretical model that consistently reproduces the measured propulsion speeds. Flow fields around the droplets are characterized by particle image velocimetry. Our results demonstrate that salt–surfactant composition governs active droplet propulsion by jointly controlling micellar solute interaction lengthscales and medium viscosity.

\end{abstract}

%\keywords{Suggested keywords}%Use showkeys class option if keyword
                              %display desired
\maketitle

%\tableofcontents
\begin{figure}[b]
\includegraphics[height=7.8cm]{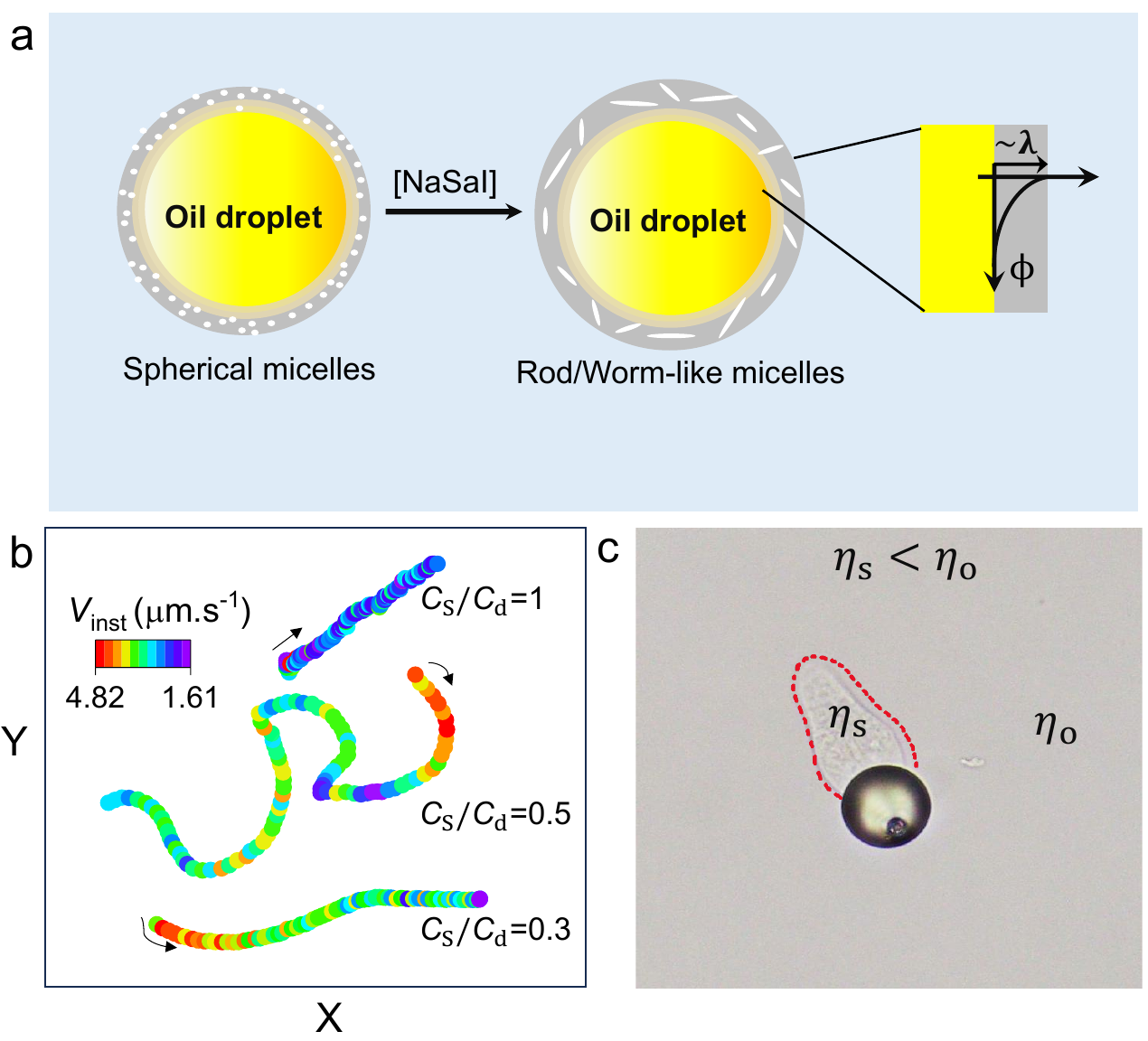}% Here is how to import EPS art
\caption{\label{fig:1} (a) Schematic representation of the variation in the interaction length scale at the droplet interface, governed by changes in the micellar structure. The droplet is positioned at a solute concentration gradient, interacting with the solute via a potential ($\phi$) confined to a diffuse layer with a thickness of several $\lambda$, where $\lambda$ denotes the solute interaction length scale. (b) Representative X-Y trajectories for active 5CB droplets with a diameter nearly identical across different cases, corresponding to concentration ratios $C_\mathrm{s}/C_\mathrm{d} = 0.3$, 0.5, and 1. The trajectories are color-coded by droplet velocity, with the direction of motion indicated by the associated arrows. (c) Optical micrograph showing a snapshot of a 5CB droplet (diameter $\sim 90~\mu$m) motion in a $C_\mathrm{s}/C_\mathrm{d} = 1$, illustrating a low-viscosity trail region (indicated by the red dotted envelope, with local viscosity $\eta_\mathrm{s}$) within a surrounding medium of viscosity $\eta_\mathrm{o}$.} 
\end{figure}

Self-propelled droplets are a simple yet robust system for studying active colloids \cite{michelin2023self,maass2024self,kumar2024emergent,thutupalli2011swarming,morozov2019nonlinear,dwivedi2023mode}. Their motion arises from interfacial chemical reactions or solute exchange processes \cite{izri2014self, izzet2020tunable,hanczyc2007fatty, krishna2025droplets}, creating concentration gradients and driving Marangoni flows \cite{suda2021straight, hanczyc2007fatty, toyota2009self, dietrich2020microscale, krishna2024dynamic, izri2014self, izzet2020tunable, cejkova2014dynamics, peddireddy2012solubilization}. This chemohydrodynamic instability \cite{michelin2023self, jin2017chemotaxis, morozov2019nonlinear, dey2022oscillatory} enables self-propulsion without geometric or chemical asymmetry \cite{dietrich2020microscale}, distinguishing them from most solid phoretic colloids \cite{anderson1989colloid, michelin2013spontaneous, moran2017phoretic, palacci2013living, golestanian2007designing, ruckenstein1981can, michelin2013spontaneous, dietrich2020microscale}. The coupling of solute transport with interfacial flow leads to complex dynamics, both for individual droplets \cite{morozov2019nonlinear,dey2022oscillatory, padhan2023activity,dwivedi2023mode,krishna2024dynamic,michelin2013spontaneous,jin2017chemotaxis} and in collective behaviors \cite{ashraf2025emergence, jin2017chemotaxis, kumar2024emergent}. Due to their simple construction and dynamic versatility, active emulsion droplets are ideal for probing active-matter physics \cite{michelin2023self, maass2024self, bowick2022symmetry, morozov2019nonlinear} and have applications in artificial life-like activities \cite{hanczyc2010chemical, krishna2025droplets, vcejkova2017droplets,jin2017chemotaxis}, including collective phenomena \cite{kumar2024emergent, palacci2013living, ashraf2025emergence, jin2017chemotaxis,thutupalli2011swarming}, and targeted delivery \cite{vcejkova2017droplets, baulin2025intelligent, kumar2024emergent, dullweber2025shape, palacci2013living, kumar2024emergent}.

The solubilization of liquid-crystal oil droplets into nanoemulsions or filled micelles creates spatial heterogeneity in micelle distribution, generating an interfacial-tension gradient \cite{michelin2023self, maass2024self, thutupalli2011swarming, peddireddy2012solubilization, izri2014self, izzet2020tunable}. Droplets of two immiscible liquids (e.g., oil–water) stabilized by surfactants can self-propel under such gradients \cite{toyota2009self, michelin2023self, hanczyc2007fatty, izzet2020tunable, izri2014self}. Asymmetries in surface tension or chemical fields induce interfacial flow, and convective surfactant transport amplifies the inhomogeneity. Motion persists when advection dominates diffusion \cite{michelin2013spontaneous, dwivedi2023mode, michelin2023self}. This directed motion, powered by continuous energy consumption, breaks time-reversal symmetry \cite{bowick2022symmetry, morozov2019nonlinear, maass2024self, michelin2023self}. The translational speed is determined by Marangoni stress and phoretic slip, reflecting internal hydrodynamics and characteristic lengthscales \cite{izri2014self, anderson1989colloid, ruckenstein1981can, golestanian2007designing, michelin2023self}. Previous studies have manipulated viscosity and Péclet number to control propulsion speed, slowing droplets by increasing the continuous-phase viscosity $\eta_o$ via polymers \cite{dwivedi2023mode}. Adding salt to surfactant solutions reduced droplet motion by lowering the contrast in critical micellar concentration (CMC) \cite{izzet2020tunable} or controlling a salt concentration gradient induced chemotactic response \cite{cejkova2014dynamics}, weakening Marangoni stress and reducing velocity.

The role of solute interaction lengthscale \cite{izri2014self} variation across micellar transitions in surfactant baths on emulsion droplet propulsion remains unexplored. We show that the non-monotonic propulsion speed of 4-cyano-4'-pentylbiphenyl (5CB) liquid crystal oil droplets in dilute aqueous cetyltrimethylammonium bromide (CTAB) surfactant solution arises from a competition between micellar growth—setting the micelle-droplet interaction length scale—and viscous drag. We present a theoretical framework incorporating the local viscosity sensed by nanoscale solutes (oil-filled micelles) in diffusiophoretic and Marangoni-driven transport \cite{anderson1989colloid, ruckenstein1981can}, providing a foundation for controlling active-droplet propulsion in viscoelastic media. By tuning the solute interaction lengthscale $\lambda$ through changes in micellar structure and morphology—from spherical to anisotropic rod-like and worm-like micelles (WLMs) \cite{dreiss2007wormlike, lam2019structural, nemoto1993dynamic, nodoushan2021effects, kim2000effects}—we show their effect in self-propulsion, while taking into account of the simultaneous enhancement of micellar media viscosity. Details of the experimental methods can be found in the Supporting Information (SI; section A).

\begin{figure}[b]
\includegraphics[height=4.7cm]{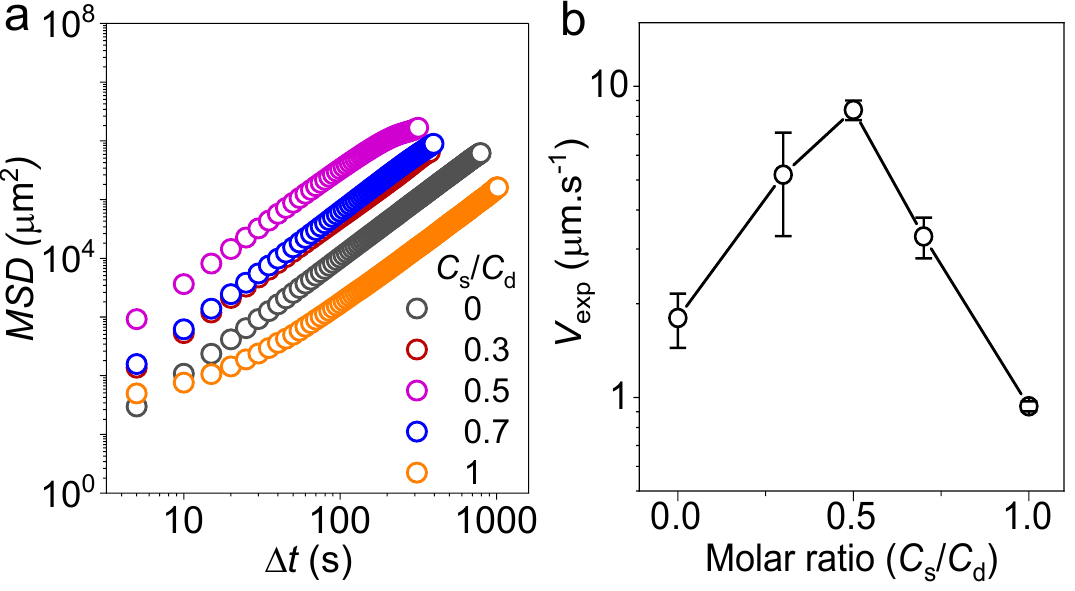}% Here is how to import EPS art
\caption{\label{fig:2} \textbf{(a)} Mean-square displacement (MSD) and \textbf{(b)} experimentally measured droplet velocity $V_{\mathrm{exp}}$, extracted from the MSD data, as functions of NaSal concentration (or equivalently $C_{\mathrm{s}}/C_{\mathrm{d}}$) at fixed CTAB concentration (10 mM).}
\end{figure}

\begin{figure*}
\includegraphics[height=5cm]{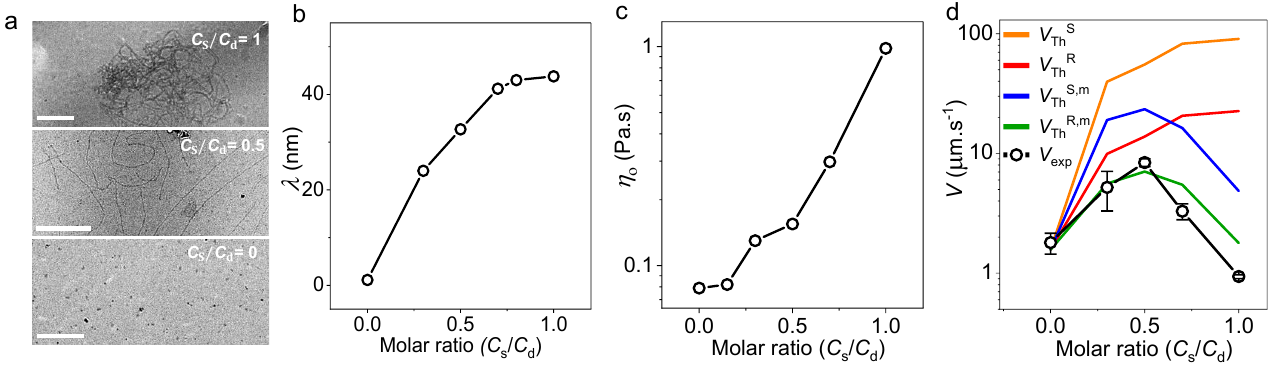}% Here is how to import EPS art
\caption{\label{fig:3} \textbf{(a)} Representative cryo-TEM images at varying NaSal concentrations, expressed as the ratio $C_{\mathrm{s}}/C_{\mathrm{d}}$ (scale bars: $xx~\mu\mathrm{m}$ in each image strip). \textbf{(b)} Solute interaction length scale $\lambda$, measured by MADLS; \textbf{(c)} zero-shear viscosity $\eta_{o}$ of the micellar medium; and \textbf{(d)} Experimentally measured droplet velocity $V_{\mathrm{exp}}$, compared with theoretical predictions of velocity considering spherical (S) solute $V^{\mathrm{S}}_{\mathrm{Th}}$ and rod-like (R) solute $V^{\mathrm{R}}_{\mathrm{Th}}$ where as $V^{\mathrm{S,m}}_{\mathrm{Th}}$ and $V^{\mathrm{R,m}}_{\mathrm{Th}}$ represent viscosity-modified theoretical velocity, as functions of NaSal concentration (or equivalently $C_{\mathrm{s}}/C_{\mathrm{d}}$). The CTAB concentration is 10 mM in all cases.}
\end{figure*}

Sodium salicylate (NaSal) salt promotes the growth of CTAB micelles, transitioning from spherical to rod-like and worm-like structures \cite{dreiss2007wormlike, lam2019structural, nodoushan2021effects, kim2000effects}. A schematic of this process is shown in Figure \ref{fig:1}a. The CTAB concentration used (10 mM) is well above its critical micelle concentration (1 mM), ensuring the presence of spherical micelles without jamming effects seen at higher concentrations. Appreciable active locomotion of 5CB droplets is observed at this low surfactant concentration (10 mM) only upon NaSal addition, an observation we encountered for the first time. Previous studies on 5CB droplet motion in CTAB have typically used higher concentrations. At 10 mM CTAB (without NaSal), 5CB droplets show only weak, short-lived motion. NaSal addition significantly enhances droplet velocity and cruising time (Figure \ref{fig:1}b).  Specifically, a $\sim 45$ {\textmu}m (radius) droplet maintained it's motion for $\sim5000$ s when 5 mm CTAB was added. 

We observe a non-monotonic dependence of droplet velocity on the NaSal-to-CTAB ratio $C_\mathrm{s}/C_\mathrm{d}$ (i.e. $C_{\mathrm{salt}}/C_{\mathrm{detergent}}$)-- propulsion speed increases, peaks near $C_\mathrm{s}/C_\mathrm{d} \approx 0.5$, and then decreases at higher NaSal concentrations. We extracted the propulsion velocity $V_\mathrm{exp}$ from fitting the MSD with the analytically derived form of an Ornstein–Uhlenbeck process (see SI, section B for details) ~\cite{uhlenbeck1930theory,das2025biophysical} and it confirms this peak–decline behavior (Figure \ref{fig:2}a-b). The viscosity of the micellar medium balances Marangoni driving stresses and viscous dissipation, governing the velocity of self-propelled droplets in the low-Reynolds-number regime. Propulsion arises from interfacial Marangoni stresses resulting from solubilization, surfactant exchange, and interfacial diffusion, while hydrodynamic resistance is determined by the viscosity of the continuous phase. Unlike prior studies, which report monotonic slowdown with viscosity- or micelle-modifying additives \cite{dwivedi2023mode, izzet2020tunable}, we observe a non-monotonic dependence of velocity on NaSal concentration.

Considering a spherical droplet of radius $a$ in a medium containing a solute of concentration $C$, droplet–solute interactions are described \cite{anderson1989colloid,izri2014self} by a potential $\phi(r)$ that depends on the distance $r$ from the interface (Figure ~\ref{fig:1}a) and arises from electrostatic, van der Waals, and steric contributions. A uniform solute distribution produces isotropic stresses and no net motion, whereas interfacial concentration asymmetry ($\nabla_{\parallel} C$) generate asymmetric tangential forces that results into self-propulsion. The resulting autophoretic velocity is  
$
V_{\mathrm{Th}} = M \langle \nabla_{\parallel} C \rangle,
$
where $M$ is the phoretic mobility and $\langle \nabla_{\parallel} C \rangle$ is the surface-averaged solute gradient \cite{anderson1989colloid,izri2014self}. Microscopically, motion in a solute gradient arises from the combined action of Marangoni stresses, characterized by the interfacial stress $\tau^\mathrm{s}$, and diffusiophoresis, described by the phoretic slip velocity $\textit{u}^\mathrm{s}$. The theoretically expected droplet speed $V_{\mathrm{Th}}$ can therefore be expressed as~\cite{ruckenstein1981can,anderson1989colloid,izri2014experimental},
\begin{eqnarray}
V_{\mathrm{Th}} =  {\frac{3\eta_\mathrm{i}}{{2\eta _\mathrm{o}}+{3 \eta_\mathrm{i}}} \langle u^\mathrm{s}\rangle} + {\frac{a}{{2\eta _\mathrm{o}}+{3 \eta_\mathrm{i}}} \langle \tau ^\mathrm{s}\rangle}.
\label{Eq1}
\end{eqnarray}
$\eta_\mathrm{o}$, $\eta_\mathrm{i}$ are the viscosity of the micellar medium and viscosity inside the 5CB oil droplet, respectively. $k_{B}$ is Boltzmann's constant. $\langle u_\mathrm{s}\rangle= \frac{k_\mathrm{B}T}{\eta_\mathrm{o}}KL^* \nabla _\parallel C$ and $\langle \tau^\mathrm{s}\rangle =k_\mathrm{B}T K\nabla _\parallel C$, represent the surface averaged slip velocity and the stress difference caused by the diffuse layer between the droplet and the background fluid~\cite{izri2014experimental}. $K$ and $L^*$ are lengthscales that characterize the distribution of solute in the diffuse layer, and related to the range and the shape of the interaction potential  ($\phi$) as $K=\int_{0}^{\infty} [exp(-\phi/ k_\mathrm{B}T) - 1]dr$,
and $L^*= K^{-1}\int_{0}^{\infty}r[exp(-\phi/ k_\mathrm{B}T)]dr$, respectively. $K$ is positive for attractive interactions and negative for repulsive interactions. The magnitude of $K$ and $L^*$ are comparable to the solute interaction length ($\lambda$) that sets the spatial extent of solute–interface coupling \cite{anderson1989colloid}. Altogether, this treatment transforms Eq.~\ref{Eq1} to

\begin{eqnarray}
V_{\mathrm{Th}}=\frac{k_\mathrm{B}T}{\eta_{o}} KL^* \left(\frac{3\eta_\mathrm{i}/\eta_\mathrm{o}+a/L^*}{2+3\eta_\mathrm{i}/\eta_{\mathrm{o}}}\right) \langle \nabla_{\parallel} C \rangle.
\label{Eq2}
\end{eqnarray}

Next, we focus on individual terms of Eq.~\ref{Eq2}. For a spherical micelle of radius (characteristic lengthscale $\lambda$), $KL^*=\lambda^{2}/2$ ~\cite{izri2014self,anderson1989colloid}. However, we derive that for a worm-like micelle $K L^*=\lambda^{2}/24$, where $\lambda$ is the characteristic length (see SI section C). We specifically consider the micelles as rigid rod of length $\lambda$ oriented arbitrarily around the sphere. Earlier theoretical work on a comparable system~\cite{izri2014self} has derived $\langle \nabla_{\parallel} C \rangle=A/D$, where $A$ quantifies the extent of activity as $A = (3/4\pi)[(da/dt)/\delta^3]$ and $D=k_{B}T/6\pi\eta_o\delta$ is the diffusivity of the oil-filled micelles following Stokes-Einstein relationship. Here $a$ is the radius of the droplet and $\delta$ is the characteristic radius of the oil-filled micelles coming out of the 5CB droplet and different from the length scale of the worm-like-micelles formed in the bulk, which we show later by experiments. Altogether, this leads to 
\begin{eqnarray}
V_{\mathrm{Th}}=k \left(\frac{da}{dt} \right)\left({\frac{\lambda}{\delta}}\right)^2 \left(\frac{3\eta_\mathrm{i}/\eta_\mathrm{o}+a\psi/\lambda}{2+3\eta_\mathrm{i}/\eta_\mathrm{o}}\right),
\end{eqnarray}
where $k=3/16$ and $\psi=6$ for worm-like (rod-shaped) micelles, and $k=9/4$ and $\psi=2$ for spherical micelles, referred as $V_{\mathrm{Th}}^{\mathrm{R}}$ and $V_{\mathrm{Th}}^{\mathrm{S}}$, respectively, further (details in SI; section D).

To obtain a theoretical estimate of the droplet velocity, the characteristic length scales of the micellar solutes, $\lambda$ and $\delta$, were determined using multi-angle dynamic light scattering (MADLS). The solute interaction length $\lambda$ was taken as the largest dimension of empty CTAB micelles, while the swollen-micelle size $\delta$ was measured in a 5CB-saturated 10 mM CTAB solution equilibrated for 3--4~h and centrifuged to remove undissolved 5CB~\cite{izri2014self}. Empty micelles exhibit characteristic sizes of $\sim 2$--3~nm, whereas swollen micelles' diameter measures $\sim 7$--8~nm, i.e., substantially larger than the core-empty spherical micelles, confirming solubilization of 5CB within the micellar core (see SI, Figure~S1a,b).

Addition of NaSal at fixed CTAB concentration promotes longitudinal micellar growth via electrostatic screening and salicylate penetration \cite{lam2019structural,dreiss2007wormlike}. The resulting decrease in zeta potential with increasing $C_\mathrm{s}$ (SI, Fig.~S1c) reflects reduced surface charge density \cite{shukla2008zeta}, driving a transition from spherical micelles to wormlike micelles (WLMs). The micellar length $\lambda$ increases from $\sim2$ nm to $\sim50$ nm and saturates for $C_\mathrm{s}/C_\mathrm{d}\simeq0.5$--$0.7$ due to charge-screening saturation (Figure~\ref{fig:3}b), whereas the zero-shear viscosity $\eta_o$ continues to increase up to $C_\mathrm{s}/C_\mathrm{d}=1.0$ (Figure~3b), indicating that viscosity enhancement is governed by micellar crowding and entanglement density rather than $\lambda$ alone \cite{dreiss2007wormlike,nodoushan2021effects,kim2000effects}. Salicylate association reduces micellar bending stiffness, yet the chains remain semiflexible and near-rodlike \cite{dreiss2007wormlike,lam2019structural}, consistent with the rod-based framework adopted in Eq.~3 (see SI, Section~C). Cryo-TEM at $C_\mathrm{s}/C_\mathrm{d}\ge0.5$ confirms long, entangled micellar networks (Figure~3a; SI, Figure~S2), where the contour length exceeds the entanglement length \cite{bhardwaj2007filament,tan2022quantitative} and matches reported $R_\mathrm{g}$ values for CTAB micelles \cite{lam2019structural}.

Substitution of experimentally measured parameters into Eq.~3 predicts a monotonic increase of the droplet velocity both for spherical and rod-like-micelles, in contrast to the experimentally observed non-monotonic behavior (Figure~3d), revealing the inadequacy of the homogeneous Newtonian model. In such media, micrometer-sized droplets and nanometer-sized solutes experience the same viscosity, $\eta_o=\eta_s$ \cite{izri2014self,izri2014experimental} (Eq.~3). %rendering $V_{\mathrm{Th}}$ viscosity-independent .
However,in entangled WLM solutions, microscopic transport decouples from macroscopic rheology: solutes diffusing in the droplet wake experience a locally reduced viscosity due to breakdown of the Stokes–Einstein continuum limit \cite{dwivedi2023mode}. For $C_\mathrm{s}/C_\mathrm{d}\simeq0.5$--$1$, the corresponding plateau modulus $\simeq0.14$--$0.21$~Pa implies entanglement mesh size of $\simeq270$--$300$~nm \cite{bhardwaj2007filament,tan2022quantitative,mohammadigoushki2016sedimentation}. Clearly, large several microns-sized 5CB droplets are immobilized by this network, where WLMs only act as fuel by sustaining interfacial tension gradients and solubilize oil into small spherical oil-filled micelles \cite{izzet2020tunable} of radius $\delta\approx3.5$~nm (MADLS; SI, Figure~S1b). Since this is well below the entanglement mesh size, these micelles diffuse through the mesh sensing a local viscosity $\eta_S$ comparable to that of a medium that forms spherical micelles~\cite{dwivedi2023mode}, rather than the viscosity $\eta_o$ of the bulk medium that forms worm-like-micelles.

Thus, the viscous drag on oil-filled micelles differs from that on the droplet, $\eta_o\neq\eta_s$, modifying the velocity expression into
\begin{eqnarray}
V^{\mathrm{m}}_\mathrm{Th}=k \left(\frac{da}{dt} \right)\left({\frac{\lambda}{\delta}}\right)^2 \left(\frac{\eta_s}{\eta_o} \right)\left(\frac{3\eta_\mathrm{i}/\eta_\mathrm{o}+a\psi/\lambda}{2+3\eta_\mathrm{i}/\eta_\mathrm{o}}\right),
\end{eqnarray}
where $k$ and $\psi$ has usual values depending the shape of the micelles. We refer to this modified droplet propulsion velocity for worm-like (rod-shaped) micelles, and spherical micelles, as $V_{\mathrm{Th}}^{\mathrm{R,m}}$ and $V_{\mathrm{Th}}^{\mathrm{S,m}}$, respectively, further (details in SI; section D).

The modified propulsion velocity accounting separately for the local viscosity of oil-filled micelles ($\eta_\mathrm{s}\sim0.08$~Pa.s) and viscosity of the bulk micellar medium ($\eta_\mathrm{o}$) reproduces the experimentally observed non-monotonic dependence when evaluated with experimentally measured parameters (Figure~\ref{fig:3}d) both for the spherical and worm like morphology. At $C_\mathrm{s}/C_\mathrm{d}=0$, micelles are predominantly spherical and $V_{\rm exp}$, $V_{\mathrm{Th}}^{\mathrm{R,m}}$ and $V_{\mathrm{Th}}^{\mathrm{S,m}}$ converge with each other. As revealed from the Cryo-TEM imaging, for $C_\mathrm{s}/C_\mathrm{d}>0$, the micelles become more worm-like and the theoretical estimate for rod-like morphology becomes significantly closer to the experimentally observed value in comparison to the theoretical estimate from sphere like morphology. Specifically,  $V_{\mathrm{Th}}^{\mathrm{S,m}}$ is almost an order of magnitude higher than $V_{\rm exp}$ suggesting the physical significance of the rod-like morphology in the theoretical analysis. Summarily, the non-monotonic trend arises from the competing contribution of the micellar interaction length $\lambda$ and the bulk viscosity $\eta_o$. $V^{\mathrm{m}}_\mathrm{Th}$ increases for $0<C_\mathrm{s}/C_\mathrm{d}<0.5$ as $\lambda$ increases. At $C_\mathrm{s}/C_\mathrm{d}>0.5$, $\lambda$ saturates and the increase in viscosity results into the reduction in propulsion velocity.
At these higher concentrations, threadlike entangled WLMs (Figure~1a, S2) constrain rod orientation, causing residual deviations between theory and experiment. These can be reconciled by assuming the rod orientation is no longer isotropic with the surface, i.e., there is a preferred angle of the micellar orientation with respect to the surface normal. This can be modeled via a von Mises–Fisher distribution whose stiffness grows with entanglement density, effectively closing the gap between theoretical expectation and experimental observation of droplet propulsion velocity (details in SI, section E and Figure~S3).

\begin{figure}[h!]
\includegraphics[height=7.2cm]{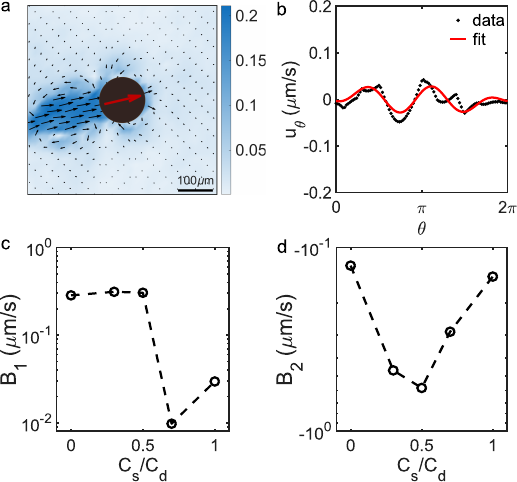}% Here is how to import EPS art
\caption{\label{fig:4} (a) Flow profile using PIV measurements at concentration $C_\mathrm{s}/C_\mathrm{d}$ = 1. (b) Tangential flow velocity, obtained from PIV measurements at $C_\mathrm{s}/C_\mathrm{d}= 1$, at the droplet interface, in the co-moving frame of reference. Red line is theoretical fits of $v_\mathrm{\theta}$. (c) $B_1$ and (d) $B_2$ value as a function of $C_\mathrm{s}/C_\mathrm{d}$.}
\end{figure}

To interpret the hydrodynamic swimming profiles of 5CB droplets as a function of NaSal concentration, Particle Image Velocimetry (PIV) was used to map flow fields (Figure \ref{fig:4}a, SI Figure S4). We fit the tangential component of the velocity field as the sum of the first two eigen modes $u_\theta (\theta)=B_1 \sin(\theta-\theta_0)+\frac{B_2}{2} \sin(2(\theta-\theta_0))$, where $\theta_0$ is the direction of propulsion with respect to the x-axis of the frame (Figure \ref{fig:4}b) ~\cite{dwivedi2023mode}.  The coefficient of the second sine mode, $B_2$, serves as a hydrodynamic fingerprint of the stresslet \cite{suda2021straight}. In our experiments, the negative sign of $B_2/B_1$ indicates a pusher-type motion of the droplet, which remains unchanged upon addition of NaSal (Figure \ref{fig:4}c). Notably, the magnitude of $B_2$ varies similar to the droplet velocity, revealing a non-monotonic dependence of the stresslet strength with a maximum at $C_\mathrm{s}/C_\mathrm{d}=0.5$ (Figure \ref{fig:4}d). The first mode, $B_1$, corresponds to the primary translational propulsion of the droplet and generally decreases with increasing NaSal, qualitatively reflecting the enhanced viscosity and micellar entanglement of the medium \cite{de2015locomotion,dwivedi2023mode}.

In summary, this study demonstrates that the self-propulsion of 5CB droplets in CTAB/NaSal media exhibits a non-monotonic velocity profile with respect to the amount of NaSal, with a peak occurring around \( C_\mathrm{s}/C_\mathrm{d} \approx 0.5 \). Our findings indicate that this behavior is governed by two competing factors resulting from the structural evolution of the propelling micellar media: the micellar solute interaction lengthscale \(\lambda \), which promotes propulsion as it increases, and micellar network entanglement, which raises the local viscosity \( \eta_o \), leading to enhancement of viscous drag. Rheological measurements confirm the onset of network entanglement at \( C_\mathrm{s}/C_\mathrm{d} \approx 0.5 \). Moreover, we show that only a theoretical model that considers worm-like geometry of the micelles and accounts for a different viscosity in the intermediate diffuse layer from the bulk entangled medium can quantitatively reproduce the experimentally observed propulsion speed as a function of NaSal concentration. Additionally, the PIV analysis supports the notion that the droplets act as a pusher in the surrounding medium. Altogether, this work demonstrates that nanoscale solute interaction length scales and local viscosity critically regulate macroscopic self-propulsion, providing a framework to rationalize experimentally observed deviations from theoretical predictions in active systems far from equilibrium. The observed nonmonotonic response offers a stringent test of theory. Future studies should elucidate the role of micellar structure beyond lengthscales, including dynamical timescales associated with wormlike micelles as "living polymers" and collective effects of multiple self-propelling droplets.

\begin{acknowledgments}
M.C. acknowledges helpful discussions with Tapomoy Bhattacharjee, Jasna Brujic, Sivasurender Chandran, and Shashi Thutupalli, and support from the Anusandhan National Research Foundation (ANRF, formerly SERB) under Grant No. CRG/2023/006329. S.D. acknowledges support from the ANRF PM ECRG No. ANRF/ECRG/2024/001067/ENS. P.P. acknowledges support from PMRF for the research studentship from MoE, India. We thank Sivasurender Chandran, Ashutosh Kumar, and Sunita Srivastava for extending support with viscosity, MADLS, and surface tension measurements, respectively. MADLS is a IIT Bombay BSBE facility and the Rheometer is funded through DST-FIST at IIT Kanpur.
\end{acknowledgments}

\bibliography{apssamp}
%\putbib% Produces the bibliography via BibTeX.

% --- End of Main Article ---

% --- Start of Supplementary Information ---
\clearpage % Start on a new page
\onecolumngrid % This is a REVTeX-specific command for single column

\begin{center}
    \textbf{\large Supplementary Information}
\end{center}

\setcounter{equation}{0}
\setcounter{figure}{0}
\setcounter{table}{0}
\setcounter{page}{1}
\makeatletter
\renewcommand{\theequation}{S\arabic{equation}}
\renewcommand{\thefigure}{S\arabic{figure}}
%\renewcommand{\bibnumfmt}[1]{[S#1]}
%\renewcommand{\citenumfont}[1]{S#1}

% Your Supplementary text goes here
%\begin{bibunit}

\subsection{Materials and Methods}
All chemicals were obtained from Loba Chemie Pvt. Ltd. (India), except 4-cyano-4’-pentylbiphenyl (5CB), which was sourced from Alfa Aesar, Thermo Fisher Scientific, Waltham, MA, USA. Aqueous solutions contained a fixed cetyltrimethylammonium bromide (CTAB) concentration (10 mM) and varying sodium salicylate (NaSal) concentrations to systematically control the salt-to-surfactant molar ratio ($C_\mathrm{s}/C_\mathrm{d}$). 5CB was dispersed at a low mass fraction (0.10 wt\%) to produce droplets in CTAB aqueous solution (20–100 µm diameter) while minimizing droplet–droplet interactions \cite{dwivedi2021solute}. Droplet motion was confined within a Hele–Shaw optical cell made from borosilicate glass slides with a 100 µm gap. Dynamics were recorded using an Olympus SZ2-ILST stereo-zoom microscope with a CCD camera, with the cell mounted on a temperature-controlled stage at 25 $^\mathrm{o}\mathrm{C}$ to prevent temperature-dependent variations in viscosity and diffusivity. To quantify the surrounding flow field, 1 µm fluorescent polystyrene tracer particles were added to the external aqueous phase, and Particle Image Velocimetry (PIV) was performed and analyzed in ImageJ and MATLAB. Multi-angle dynamic angle scattering (MADLS, Wyatt Technology) was employed to characterize the surfactant media at different NaSal concentrations by measuring the micellar diffusion coefficient, which provides a direct estimate of the average micelle size in dilute solution. Rheological measurements were performed using a rheometer (MCR 702e Space, Anton Paar, Austria). All experiments were conducted using a cone–plate (CP) geometry with a cone diameter of 25 mm and a cone angle of $2\,^{\mathrm{o}}$. All measurements were taken at a controlled room temperature of $25\, ^{o}$C. Zeta potential  of the surfactant media is measured in Malvern Zetasizer ZS DLS instrument that reveals the surface potential of the miceller structure.

\subsection{Deriving droplet propulsion velocity from MSD}

The active force in our model represents the motility and is inspired by the Ornstein-Uhlenbeck process~\cite{sevilla2019generalized,uhlenbeck1930theory}. Active force ($\vec F_i^{\rm A}$) on particle $i$ evolves as,
\begin{equation}
    d\vec{F}^{\rm A}_{i}=-k_{a}\vec{F}^{\rm A}_{i} dt +\Gamma d\vec{W}_{t},
    \label{OUProcess}
\end{equation}
where, ${1}/{\hat{k}_{a}}$ represents the timescale of persistence of the  motion, and $d\vec{W}_{\hat{t}}$ represents a Gaussian white noise, and $\hat \Gamma$ represents the strength of motility. Following the properties of an Ornstein-Uhlenbeck process, the time correlation of the force evolves as~\cite{das2025biophysical},
\begin{equation}
    \langle \vec{F}^{\rm A}_i(s) \cdot \vec{F}^{\rm A}_i(s')\rangle=\frac{d\Gamma^2}{k_a }e^{-k_a|s-s'|}, \quad s,s'\geq0
    \label{Correlation},
\end{equation}

Now,
\begin{equation}
    \frac{\partial \vec x_i}{\partial t} =\frac{1}{\eta} \vec{F_i^{\rm A}} \quad \Longrightarrow \Delta x=\int_0^t \frac{1}{\eta} F_i^{\rm A}dt 
    \label{Delta_x},
\end{equation}
where, $\eta$ is the viscosity of the medium.

The Mean Squared Displacement can be derived by taking an ensemble average of the equation~\ref{Delta_x}, which becomes:
\begin{equation}
    \langle\Delta x^2\rangle=\frac{1}{\eta^2}\int_0^tds\int_0^t F_i^{\rm A}(s)F_i^{\rm A}(s')ds' \quad\Longrightarrow \langle\Delta x^2\rangle=\frac{d\Gamma^2}{\eta^2k_a}\int_0^tds\int_0^t e^{-k_a|s-s'|}ds'
\end{equation}

Evaluating the double integral using symmetry:
\begin{equation}
    \int_0^tds\int_0^tds'e^{-k_a|s-s'|}=2 \int_0^tds(t-s)e^{-k_as}
\end{equation}

To evaluate the double integral
\[
I=\int_0^t\!\mathrm{d}s \int_0^t\!\mathrm{d}s'\, e^{-k_a|s-s'|}
\]
we use the symmetry of the integrand in the square domain \([0,t]\times[0,t]\). We split the region along the diagonal \(s'=s\), separating the cases \(s'>s\) and \(s'<s\):
\[
\begin{aligned}
I
&= \int_0^t\!\mathrm{d}s
   \left[
      \int_0^{s}\!\mathrm{d}s'\, e^{-k_a(s-s')}
      + \int_{s}^{t}\!\mathrm{d}s'\, e^{-k_a(s'-s)}
   \right]
\end{aligned}
\]

Introducing the change of variables \(u = s-s'\) in the first integral and \(u = s'-s\) in the second yields
\[
I = \int_0^t\!\mathrm{d}s
    \left[
        \int_0^{s}\!\mathrm{d}u\, e^{-k_a u}
        + \int_0^{\,t-s}\!\mathrm{d}u\, e^{-k_a u}
    \right]
\]

The two terms are identical upon exchanging \(s\leftrightarrow t-s\), so the expression can be written as
\[
I = 2\int_0^t\!\mathrm{d}s \int_0^{\,t-s}\!\mathrm{d}u\, e^{-k_a u}
\]
Evaluating the inner integral,
\[
\int_0^{\,t-s} e^{-k_a u}\,\mathrm{d}u
   = \frac{1 - e^{-k_a(t-s)}}{k_a},
\]
gives
\[
I = \frac{2}{k_a}\int_0^t\!\mathrm{d}s\,
       \big[1 - e^{-k_a(t-s)}\big]
\]
Integrating by parts,
\[
I = 2\int_0^t (t-s)\, e^{-k_a s}\, \mathrm{d}s
\]
\[
I = \frac{2}{k_a^2}\,\big(k_a t - 1 + e^{-k_a t}\big)
\]

Therefore, the MSD is,
\begin{equation}
    \langle\Delta x^2(t)\rangle=\frac{2d\Gamma^2}{\eta^2k_a^3}(k_at-1+e^{-k_at})
\end{equation}
where $d$ is the dimension of the simulation box.

From the equation of motion, the ensemble-averaged instantaneous velocity $\vec v_i(\rm t)$ 
\begin{equation}
      \vec v_i(\rm t)=\frac{\partial \vec x_i}{\partial t} =\frac{1}{\eta} \vec{F_i^{\rm A}}, 
  \label{NonD_EoM}
\end{equation}
The velocity-velocity correlation evolves as:
\begin{equation}
    \langle \vec{v}^{\rm A}_i(s) \cdot \vec{v}^{\rm A}_i(s')\rangle=\frac{d\Gamma^2}{\eta^2 k_a}e^{-k_a|s-s'|}
    \label{Correlation},
\end{equation}
% Define time lag
The time lag can be defined as, 
\[\Delta t \equiv |s-s'|\]
% Velocity autocorrelation (given)
Therefore, the velocity-velocity correlation can be written as:
\begin{equation}
\langle \vec{v}^{\rm A}_i(s)\cdot\vec{v}^{\rm A}_i(s')\rangle
=\frac{d\Gamma^2}{\eta^2 k_a}\,e^{-k_a\Delta t}
\equiv \langle v_i^2\rangle\,e^{-k_a\Delta t},
\qquad\text{\rm where}\quad
\langle v_i^2\rangle=\frac{d\Gamma^2}{\eta^2 k_a}.
\end{equation}
Expanding the exponential term $e^{-k_a\Delta t}$,
\[
e^{-k_a\Delta t}
\approx 1 - k_a\Delta t + \frac{1}{2}(k_a\Delta t)^2+\mathcal{O}(k_a\Delta t)^3
\]
At lower timescales, the means-squared displacement shows a ballistic behavior,
\[
\langle\Delta x^2(t)\rangle=\frac{d\Gamma^2}{\eta^2k_a}t^2=\langle v_i^2\rangle t^2
\]
However, at higher timescales, the exponential term becomes negligible,
\[
e^{-k_a\Delta t}
\approx 0
\]
And, the MSD shows a linear relationship with t.
\[
\langle\Delta x^2(t)\rangle=\frac{2d\Gamma^2}{\eta^2k_a^2}(t-\frac{1}{k_a})\approx2D t, \quad {\rm where}, \quad D=\frac{d\Gamma^2}{\eta^2k_a^2}.
\]

$v_\mathrm{i}$ is extracted from the fitting and represented as $V_\mathrm{exp}$ in the main manuscript.

\subsection{Derivation of $K$ and $L^*$ from Anderson's definition with Gouy--Chapman hypothetical concentration profile of solute near a
solid surface with hard potential}\label{rod_derivn}

We begin from the Anderson/Mayer-like definition \cite{anderson1989colloid}
\[
f(y)\;=\;e^{-\phi(y)/k_{\mathrm{B}}T}-1,
\qquad
K\;=\; -\Big\langle \int_{0}^{\infty} f(y)\,dy \Big\rangle_{\theta},
\]
where, $f(y)$ represents the Gouy--Chapman concentration profile from the surface of the droplet, \(\phi(y)\) is the interaction potential as a function of the normal coordinate \(y\) measured from the surface and \(\langle\cdot\rangle_{\theta}\) denotes averaging over rod orientations. The minus sign is chosen so that a purely excluded-volume (hard potential) contribution gives a positive \(K\).

For a hard (infinite) potential produced by a rod of length \(\ell\) oriented at polar angle \(\theta\) (angle between rod axis and surface normal), the center-of-rod is excluded from the region \(0\le y < y_{\mathrm{cut}}(\theta)\) with
\[
y_{\mathrm{cut}}(\theta)=\frac{\ell\cos\theta}{2},
\]
because the rod (centered at height \(y\)) would overlap the surface if the center lies closer than half the projected length. Previous studies have noted that the persistence length of the micelles in our system are much higher than the lengths of the rods we see in our system, and can hence be treated as stiff rods \cite{wormlikemicelle_rheology}. Thus the potential may be represented as
\[
\phi(y,\theta)=
\begin{cases}
+\infty, & 0\le y < \dfrac{\ell\cos\theta}{2},\\[6pt]
0, & y \ge \dfrac{\ell\cos\theta}{2}.
\end{cases}
\]

The Mayer function \(f(y,\theta)=e^{-\phi/k_{\mathrm{B}}T}-1\) is therefore
\[
f(y,\theta)=
\begin{cases}
-1, & 0\le y < \dfrac{\ell\cos\theta}{2},\\[4pt]
0, & y \ge \dfrac{\ell\cos\theta}{2}.
\end{cases}
\]

Integrating in \(y\) gives the orientation-dependent integral
\[
\int_{0}^{\infty} f(y,\theta)\,dy
= \int_{0}^{\ell\cos\theta/2}(-1)\,dy
= -\,\frac{\ell\cos\theta}{2}.
\]

We now average over orientations restricted to the hemisphere \(0\le\theta\le\pi/2\). The normalized hemispherical distribution for an isotropic orientation on the allowed hemisphere is
\[
p(\theta)=\sin\theta,\qquad 0\le\theta\le\frac{\pi}{2},
\quad\text{with}\quad \int_{0}^{\pi/2}\sin\theta\,d\theta=1.
\]
Thus the averaged integral is
\[
\Big\langle \int_{0}^{\infty} f(y,\theta)\,dy \Big\rangle_{\theta}
= \int_{0}^{\pi/2}\left(-\frac{\ell\cos\theta}{2}\right)\sin\theta\,d\theta
= -\frac{\ell}{2}\int_{0}^{\pi/2}\cos\theta\sin\theta\,d\theta.
\]

Hence
\[
\Big\langle \int_{0}^{\infty} f(y,\theta)\,dy \Big\rangle_{\theta}
= -\frac{\ell}{2}\cdot\frac{1}{2}
= -\frac{\ell}{4}.
\]

Finally, recalling the sign convention chosen for \(K\),
\[
K \;=\; -\Big\langle \int_{0}^{\infty} f(y,\theta)\,dy \Big\rangle_{\theta}
\;=\; -\left(-\frac{\ell}{4}\right)
\;=\; \frac{\ell}{4}.
\]

Similarly, the definition of $KL^*$ as the first moment of $f(y,\theta)$ would imply that :
\begin{equation}
    KL^* = \Big\langle \int y f(y) dy \Big\rangle_{\theta} = \frac{l^2}{24}
\end{equation}
Hence we can deduce the value of $L^*$ as $\frac{l}{6}$

For a neutral solute modeled as a hard sphere, the length scale \textit{l} represents the radius of the solute, and the solute--surface interaction is purely steric. The potential of mean force is therefore
\[
\phi(y)=
\begin{cases}
\infty, & 0<y<l,\\
0, & y>l,
\end{cases}
\]
where \(y\) is the distance of the solute center from the surface. Consequently,
\(e^{-\phi(y)/k_BT}-1=-1\) for \(0<y<l\) and vanishes for \(y>l\).
Using definitions from \cite{anderson1989colloid},
\[
K=-\int_0^\infty \bigl(e^{-\phi(y)/k_BT}-1\bigr)\,dy,
\qquad
K L^*=\int_0^\infty y\,\bigl(e^{-\phi(y)/k_BT}-1\bigr)\,dy,
\]
the integrals reduce to
\[
K=\int_0^l (1)\,dy=l,
\qquad
K L^*=\int_0^l (-y)\,dy=-\frac{l^2}{2}.
\]
Thus, for a hard-sphere solute one obtains
\[
K=l,
\qquad
L^*=\frac{l}{2},
\]
and the phoretic mobility is proportional to the geometric moment
\(K L^*=-l^2/2\), reflecting the thickness and centroid of the solute depletion layer adjacent to the surface.

\begin{figure}
\centering
    \includegraphics[width= 1\linewidth]{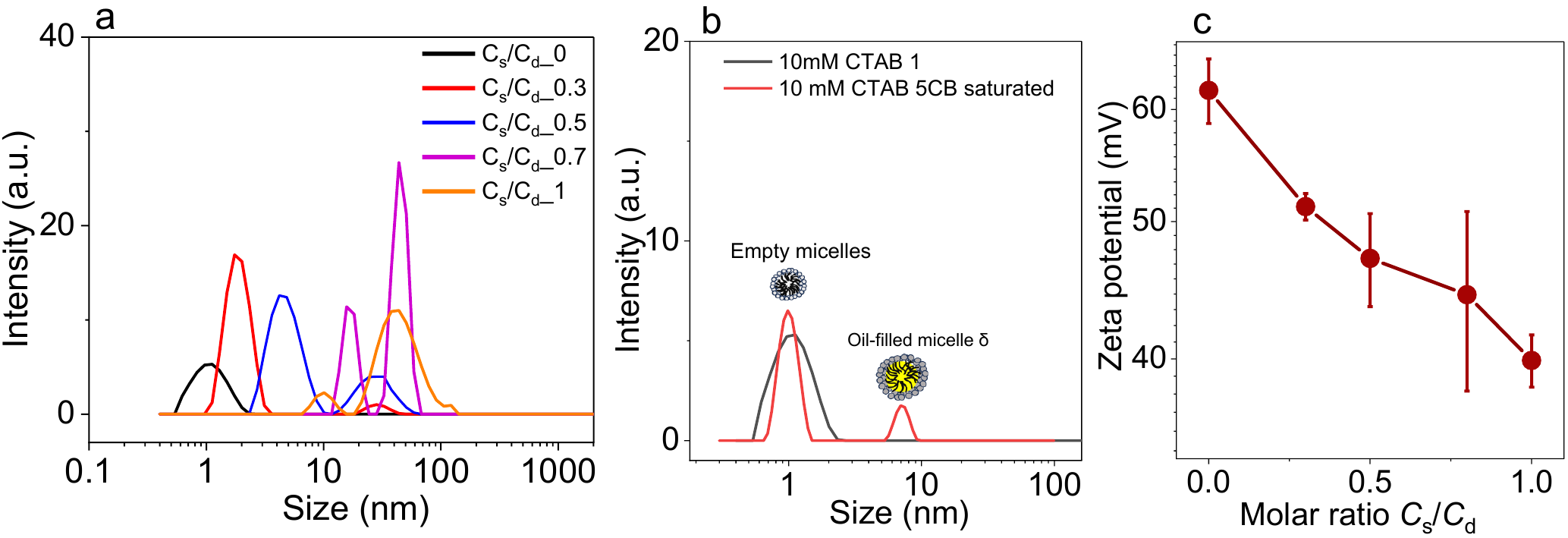}
    \caption{(a) MADLS data with variation of $C_\mathrm{s}/C_\mathrm{d}$ (b) MADLS data of 10 mM CTAB and 5CB oil saturated 10 mM CTAB containing dimension ($\delta$) of oil filled micelles. (c) Zeta potential of  10 mM CTAB with variation of NaSaI salt
    ($C_\mathrm{s}/C_\mathrm{d}$). }
    \label{S1}
\end{figure}

\subsection{Theoretical Velocity Calculation}
The theoretical liquid droplet velocity ($V_\mathrm{Th}$)
 \begin{eqnarray}
V_{\mathrm{Th}}=\frac{k_\mathrm{B}T}{\eta_{o}} KL^* \left(\frac{3\eta_\mathrm{i}/\eta_\mathrm{o}+a/L^*}{2+3\eta_\mathrm{i}/\eta_{\mathrm{o}}}\right) \langle \nabla_{\parallel} C \rangle
\end{eqnarray}
Considering the $\langle \nabla_{\parallel} C \rangle$ term the equation becomes
 \begin{eqnarray}
V_{\mathrm{Th}}=\frac{9}{2} \left(\frac{da}{dt}\right) \frac{\eta_\mathrm{s}}{\eta_\mathrm{o}} \frac{KL^*}{\delta^2}  \left(\frac{3\eta_\mathrm{i}/\eta_\mathrm{o}+a/L^*}{2+3\eta_\mathrm{i}/\eta_{\mathrm{o}}}\right) 
\end{eqnarray}
System-1: In case $\eta_\mathrm{o}=\eta_\mathrm{s}$ and  considering a spherical solute, ($KL^*=\lambda^2/2$) the velocity represented as ($V^\mathrm{S}_{\mathrm{Th}}$) and for rod-like solute  ($KL^*=\lambda^2/24$), the velocity written as ($V^\mathrm{R}_{\mathrm{Th}}$). The corresponding equations are
 \begin{eqnarray}
V^\mathrm{S}_{\mathrm{Th}}=\frac{9}{4} \left(\frac{da}{dt}\right)  \frac{\lambda^2}{\delta^2}  \left(\frac{3\eta_\mathrm{i}/\eta_\mathrm{o}+2a/\lambda}{2+3\eta_\mathrm{i}/\eta_{\mathrm{o}}}\right) 
\end{eqnarray}

\begin{eqnarray}
V^\mathrm{R}_{\mathrm{Th}}=\frac{3}{16} \left(\frac{da}{dt}\right)  \frac{\lambda^2}{\delta^2}  \left(\frac{3\eta_\mathrm{i}/\eta_\mathrm{o}+ 6a/\lambda}{2+3\eta_\mathrm{i}/\eta_{\mathrm{o}}}\right) 
\end{eqnarray}
System-2: In case $\eta_\mathrm{o}\neq\eta_\mathrm{s}$ and  considering a spherical solute, ($KL^*=\lambda^2/2$) the viscosity modified velocity represented as ($V^\mathrm{S,m}_{\mathrm{Th}}$) and for rod-like solute  ($KL^*=\lambda^2/24$), the velocity written as ($V^{\mathrm{R,m}}_{\mathrm{Th}}$). The corresponding equations are
 \begin{eqnarray}
V^{\mathrm{S,m}}_{\mathrm{Th}}=\frac{9}{4} \left(\frac{da}{dt}\right)\frac{\eta_\mathrm{s}}{\eta_\mathrm{o}}  \frac{\lambda^2}{\delta^2}  \left(\frac{3\eta_\mathrm{i}/\eta_\mathrm{o}+2a/\lambda}{2+3\eta_\mathrm{i}/\eta_{\mathrm{o}}}\right) 
\end{eqnarray}
\begin{eqnarray}
V^{\mathrm{R,m}}_{\mathrm{Th}}=\frac{3}{16} \left(\frac{da}{dt}\right) \frac{\eta_\mathrm{s}}{\eta_\mathrm{o}} \frac{\lambda^2}{\delta^2}  \left(\frac{3\eta_\mathrm{i}/\eta_\mathrm{o}+ 6a/\lambda}{2+3\eta_\mathrm{i}/\eta_{\mathrm{o}}}\right) 
\end{eqnarray}
At $C_\mathrm{s}/C_\mathrm{d}$=0, the micelles are mostly spherical, so at this concentration, all velocities are calculated by using the spherical model in all 4 cases. Beyond $C_\mathrm{s}/C_\mathrm{d}$=0, the rod-like solute treatment was computed in Equ. S15 and S17. Whereas in system-2, the $\eta_\mathrm{s}$ is considered as constant (0.079 Pa.s), local viscosity is considered during the theoretical calculation.

\begin{figure}
\centering
    \includegraphics[width= 0.9\linewidth]{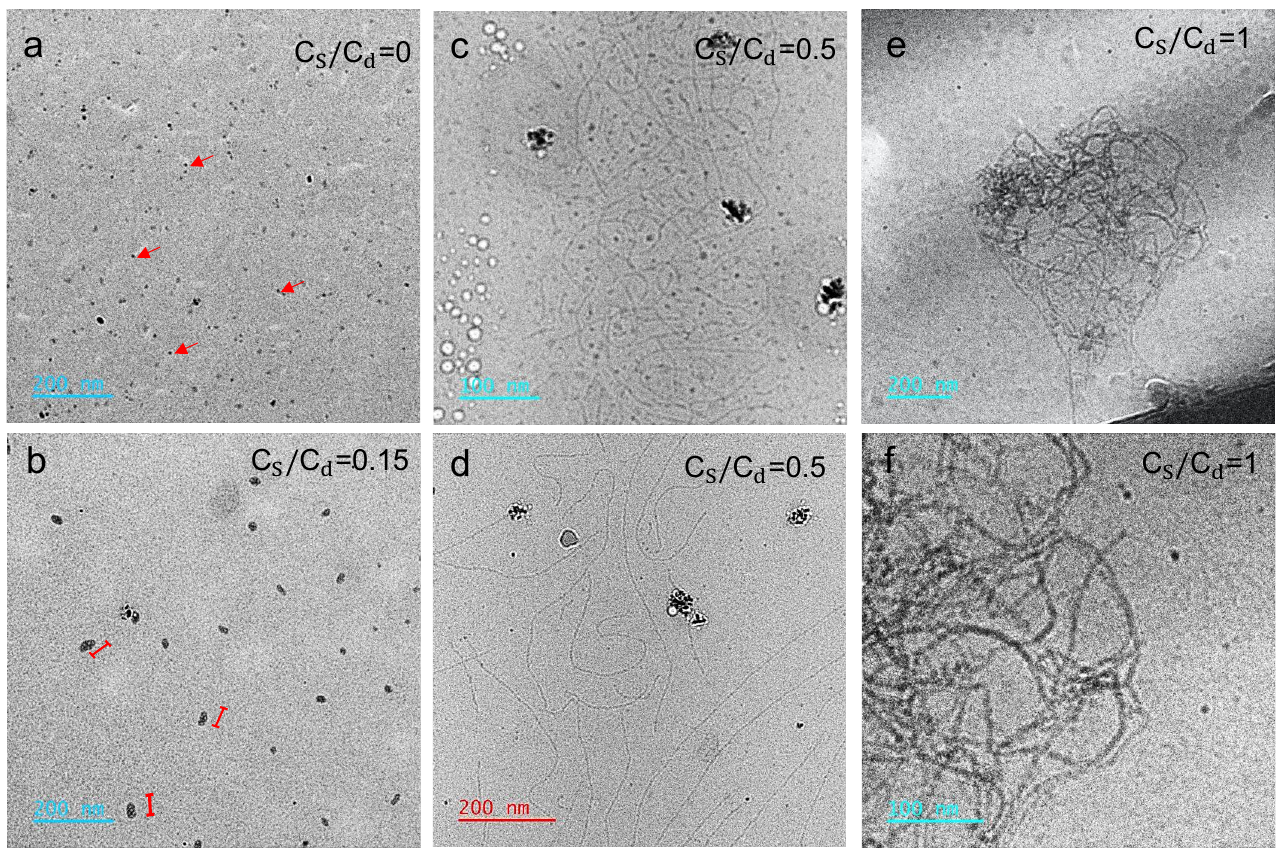}
    \caption{Cryo TEM image with variation of additive NaSal salt at constant 10 mM CTAB.(a) $C_\mathrm{s}/C_\mathrm{d}$=0, (b) $C_\mathrm{s}/C_\mathrm{d}$=0.15, (c-d) $C_\mathrm{s}/C_\mathrm{d}$=0.5, (e-f) $C_\mathrm{s}/C_\mathrm{d}$=1,  Scale bar mentioned inside.  }
    \label{S1}
\end{figure}

\begin{figure}
\centering
    \includegraphics[width= 0.6\linewidth]{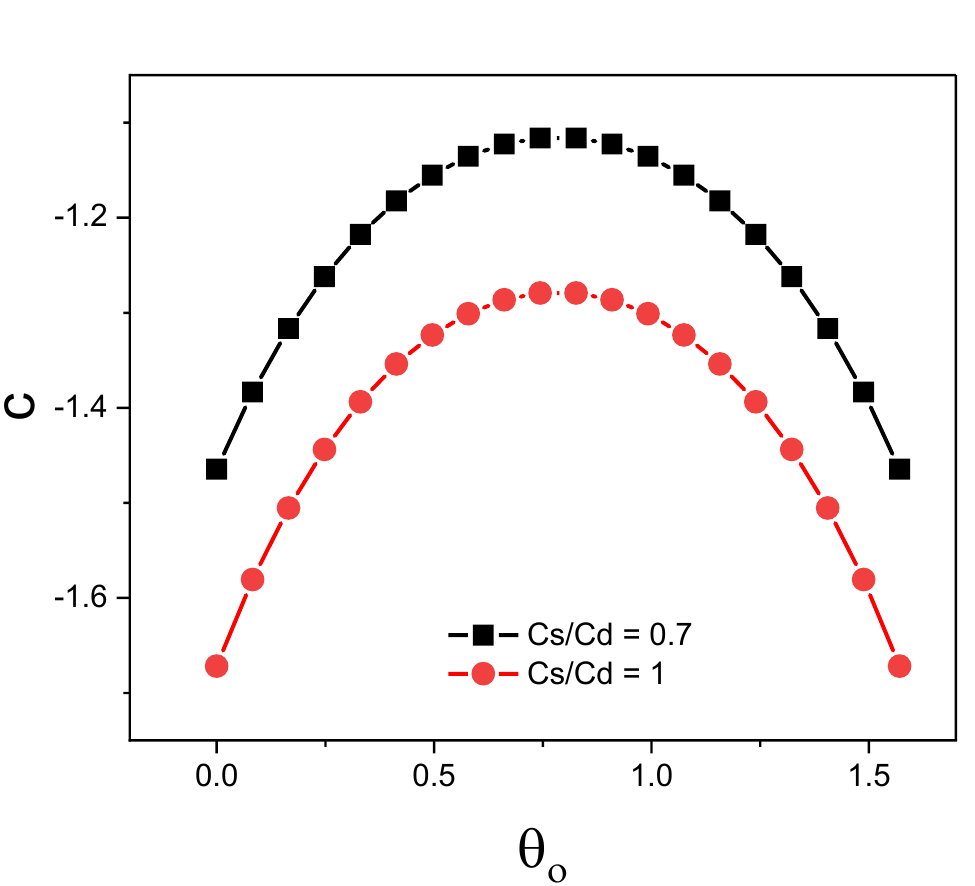}
    \caption{The numerically estimated values of the stiffness constant of the entangled media for various possible values of the preferred contact angle with the droplet surface}
    \label{c_theta}
\end{figure}

\begin{table}[h]
\centering
\caption{Parameters for theoretical velocity calculation}
\begin{tabular}{|c|c|c|c|} 
 \hline
$C_\mathrm{s}/C_\mathrm{d} $ & Interaction length, $\lambda$  $(\mathrm{\mu}m)$ & Zero-shear viscosity,$\eta_\mathrm{o}$ (Pa.s) & Droplet shrinkage rate, $dr/dt$ $(\mu m/s)$  \\ [0.5ex]

\hline
 0 &0.00112 & 0.079 & 0.01477 \\ 
 
 0.3 & 0.024 & 0.132 & 0.02088\\ 

0.5 & 0.0327 &0.155 & 0.02174 \\ 

 0.7 & 0.0438 & 0.298 & 0.01639 \\ 
 
1 & 0.0438 & 0.983 & 0.01353 \\ [0.5ex] 
\hline
 \end{tabular}
\label{table:data}
\end{table}

\subsection{Stiffness and rod orientation}
Section \ref{rod_derivn} details the derivation of $K$ and $L^*$ when the rod orientation tends to be isotropic. When this is not the case, i.e., when there is a preference in the orientation angle, the angle distribution would vary centered around a mean value $\theta_o$ and can be expressed using the normalised von Mises-Fisher (vMF) distribution given by 
\begin{equation}
    p(\theta) = \frac{c\sin\theta e^{c.\cos(\theta-\theta_o)}}{\sinh (c)}
\end{equation}
where $c$ is the stiffness constant of the entangled micellar media. It can be noted that when $ |c|\xrightarrow{}0$ this probability distribution will tend to the isotropic distribution described in the previous section. When $K$ and $L^*$ are re-derived using this probability distribution, a value of $c$ can be back calculated that can accommodate the quantitative difference between the theoretically calculated and experimental velocity. Fig. \ref{c_theta} is a plot of the various values $c$ can take for different preferred orientation angles $\theta_o$ of the entangled micelles on the droplet surface.

\begin{figure}
\centering
    \includegraphics[width= 0.8\linewidth]{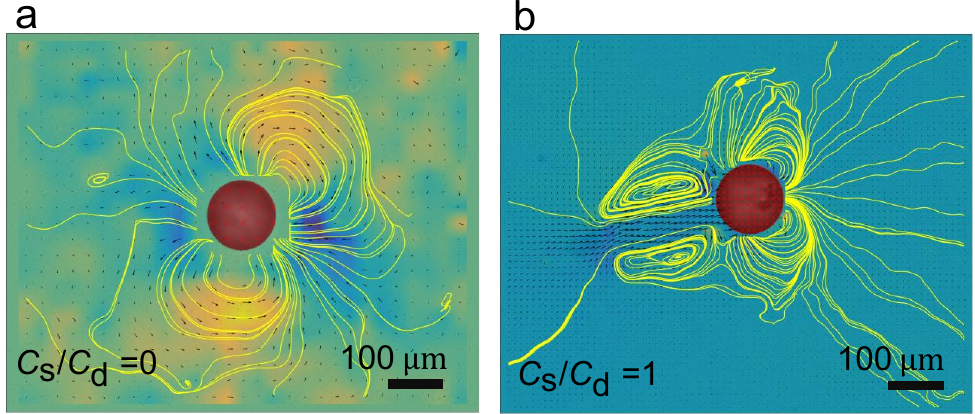}
    \caption{ (a and b) Micrographs representing streamlines obtained from PIV experiments. This streamlines represent the flow-field around the droplet. }
    \label{S1}
\end{figure}

%*\subsection*{References}
%\putbib% Produces the bibliography via BibTeX. 

%\bibliography{apssamp}

\twocolumngrid % If you need to switch back, though usually not needed at the end

%\end{bibunit}
\end{document}